# $^{12}$CO and $^{13}$CO observation of the low-metallicity dwarf galaxy DDO 154

Shinya Komugi,[1,*] Miku Inaba,[2,†] and Tetsuo Shindou[1]

[1]Division of Liberal Arts, Kogakuin University, 2665-1 Nakano-cho, Hachioji, Tokyo 192-0015, Japan
[2]Department of Applied Physics, Kogakuin University, 2665-1 Nakano-cho, Hachioji, Tokyo 192-0015, Japan

*E-mail: skomugi@cc.kogakuin.ac.jp

†Present address: Knox Data Corp., 6-20-14 Minami-oi, Shinagawa-ku, Tokyo 140-0013, Japan

## Abstract

The conversion factor from carbon monoxide (CO) intensity to molecular gas mass is a source of large uncertainty in understanding gas and its relation to star formation in galaxies. In particular, the conversion factor in low-metallicity environments have remained elusive, as currently only two galaxies have been detected in any CO isotopes in environments with $12 + \log(\mathrm{O/H}) < 8.0$. Here we report $^{12}$CO ($J = 1$–0) and $^{13}$CO ($J = 1$–0) observations towards a star-forming region in DDO 154, a low-metallicity dwarf irregular galaxy at $12 + \log(\mathrm{O/H}) = 7.67$. This is a re-observation of a previous non-detection at higher angular and velocity resolution. No significant emission was detected. By estimating the molecular gas mass from associated star formation, we find that DDO 154 has a conversion factor of more than $10^3$ times the Milky Way. Alternatively, if we estimate molecular mass using dust continuum emission, the conversion factor is at least 2 orders of magnitude larger than the Milky Way. These estimates signify a large amount of CO-dark molecular gas in this galaxy.

**Key words:**  galaxies: dwarf — galaxies: ISM — ISM: abundances

## 1 Introduction

Current theories of star formation require deposits of molecular gas prior to its triggering. Carbon monoxide (CO) has been used routinely as a tracer of molecular gas, because molecular hydrogen lacks a dipole moment and can emit only under extreme conditions. The ratio of molecular gas column density to CO intensity, i.e., the conversion factor $X_{\mathrm{CO}}$, is known to be approximately constant in the Milky Way (MW) at $X_{\mathrm{CO,MW}} = 2.0 \times 10^{20}$ [cm$^{-2}$ (K km s$^{-1}$)$^{-1}$] (Bolatto et al. 2013). At low metallicities, higher $X_{\mathrm{CO}}$ is expected, because the hard UV photons in such environments penetrate into the interstellar medium (ISM) and photodissociate CO molecules. This gives rise to molecular gas that cannot be traced by CO (i.e., "CO dark" gas), at the same time making the detection of CO in low-metallicity environments challenging. Observational constraints on the metallicity dependence of $X_{\mathrm{CO}}$ has been controversial, however. While early studies (Wilson 1995; Arimoto et al. 1996) indeed find increasing $X_{\mathrm{CO}}$ at low metallicity, others find only very weak (Bolatto et al. 2003; Sandstrom et al. 2013; Jiao et al. 2021) or no (Rosolowsky et al. 2003; Bolatto et al. 2008) variation in $X_{\mathrm{CO}}$ over a range of metallicities. The disagreements in these studies arise from the fact that most studies are based on observations of galaxies with relatively high metallicity at $12 + \log[\mathrm{O/H}] \geq 8.0$ (where the MW metallicity is 8.9).

There have been numerous attempts at detecting CO at metallicities as low as $12 + \log[\text{O/H}] < 8.0$ (Verter & Hodge 1995; Buyle et al. 2006; Komugi et al. 2011b and references therein), but only two galaxies, WLM at $12 + \log[\text{O/H}] = 7.8$ (Rubio et al. 2015) and Sextans B at $12 + \log[\text{O/H}] = 7.7$ (Shi et al. 2016, 2020) have been successful. For both galaxies, the studies find that molecular clouds are systematically smaller than MW clouds, their radii ranging from 1 to 4 pcs, but roughly follow the same size–line width relation (Solomon et al. 1987). The small CO emitting area is consistent with the interpretation that CO is photodissociated in the outer layers of clouds, resulting in CO-dark molecular gas. The conversion factor in these systems determined by assuming that the clouds are virialized, and comparing the dynamical and luminous mass, is estimated to be 7 to 30 times larger than the MW for Sextans B, and 6 times larger than the MW for WLM. When the dust continuum is used to estimate the "true" molecular gas by assuming a gas-to-dust ratio, the conversion factor in WLM is much larger, at 30 times the MW value (Elmegreen et al. 2013). The larger conversion factor determined when using dust as a proxy for gas indicates that a significant amount of molecular gas resides in regions outside where CO is detected.

An increasing number of high-redshift galaxies (thus hosting low-metallicity gas) are being detected with the advent of recent telescopes, while CO remains to be the line of choice for tracing molecular gas. Many recent studies use various tracers to cross-calibrate molecular gas and determine $X_{\text{CO}}$, such as [C I] ($^3P_1 - {}^3P_0$) and thermal dust continuum in addition to CO (Leroy et al. 2011; Bolatto et al. 2013; Sandstrom et al. 2013; Jiao et al. 2021; Dunne et al. 2022), but connecting these new tracer candidates to previous studies conducted with CO require that we calibrate $X_{\text{CO}}$ in low metallicity well.

In this study we report CO observations toward the dwarf irregular galaxy DDO 154 at a distance of 3.2 Mpc (Carignan & Purton 1998). At this distance, $1''$ corresponds to 15.5 pc. Its metallicity of $12 + \log[\text{O/H}] = 7.67$ is similar to that of WLM and Sextans B. The difference between DDO 154, WLM, and Sextans B is in their star formation properties. In WLM, the total star formation rate (SFR) and stellar mass ($M_*$) are SFR $= 6 \times 10^{-3} M_\odot$ yr$^{-1}$ and $M_* = 1.6 \times 10^7 M_\odot$ (Hunter et al. 2010; Zhang et al. 2012), respectively. The SFR per unit stellar mass (specific SFR) is therefore $4 \times 10^{-10}$ yr$^{-1}$. This is roughly 10 times the value in the MW, where SFR $= 1.9 M_\odot$ yr$^{-1}$ and $M_* = 6.4 \times 10^{10} M_\odot$ (Chomiuk & Povich 2011; McMillan 2011). Similarly, Sextans B has SFR $= 2$–$4 \times 10^{-3} M_\odot$ yr$^{-1}$ and $M_* = 4.4 \times 10^7 M_\odot$ (Weisz et al. 2011). The specific SFR of Sextans B is close to the MW value, but the star formation history derived from color–magnitude diagrams of resolved stellar populations (Weisz et al. 2011) reveal that the current SFR of Sextans B is about 2 times its lifetime average. These indicate that WLM and Sextans B are in a bursty phase of its lifetime, currently in the process of enriching the ISM with heavy elements. DDO 154 is, however, quiescent; its global SFR and stellar mass are SFR $= 1.5 \times 10^{-3} M_\odot$ yr$^{-1}$ and $M_* = 3.8 \times 10^7 M_\odot$, respectively (Carignan & Purton 1998), making its specific SFR comparable to that of the MW. Its star formation is currently 1/2 to 1/3 of its lifetime average (Kennicutt & Skillman 2001). It is also one of the most gas rich galaxies known, with $M_{\text{H\,I}} = 2.5 \times 10^8 M_\odot$, and stars accounting for only 2% of its dynamical mass (Carignan & Purton 1998). Thus, molecular gas is expected to be in a pristine environment before any major evolution has occurred in the dwarf galaxy's lifetime. Calibrating $X_{\text{CO}}$ in DDO 154 can function as a reference for studying gas in the most unevolved systems.

DDO 154 was previously observed at the $^{12}$CO ($J = 1$–0) frequency toward Region 2 (Komugi et al. 2011b; hereafter K11), the third most luminous H II region, with O/H spectroscopy by Kennicutt and Skillman (2001). The Hubble Space Telescope (HST) Advanced Camera for Surveys (ACS) $F606W$ shows a young stellar cluster associated with this H II region (figure 1). It is the only H II region within the stellar disk detected by Spitzer MIPS 24 $\mu$m (Kennicutt et al. 2003) and by Herschel SPIRE 250 $\mu$m (Kennicutt et al. 2011). No significant CO detection was made by K11, but the spectra from the IRAM 30 m telescope at $22''$ resolution showed a speculative $2.3\sigma$ signature in a single 3.3 km s$^{-1}$ width channel at a velocity of $V_{\text{lsr}} = 367$ km s$^{-1}$, which is consistent with the galaxy's global recession velocity. We conducted follow-up observations at moderately better angular resolution including the $^{13}$CO ($J = 1$–0) line, to test whether this signature is a real emission.

## 2 Observation and results

Observations were conducted at the Nobeyama Radio Observatory (NRO) 45 m telescope over 7 nights in 2020 December. Beams 1 and 3 of the four-beam dual polarization receiver FOREST (Minamidani et al. 2016) were used in the ON–ON position switching mode with 20 second integrations (i.e., one beam observing the OFF position while the other is at the ON position). Thus, a single beam was always pointed to Region 2 at (RA, Dec) $= (12^{\text{h}}54^{\text{m}}03{.}^{\text{s}}9, +27°09'05'')$ (J2000.0). The backend was the SAM45 spectrometer (Kuno et al. 2011; Kamazaki et al. 2012), with 16 spectral windows, each with 125 MHz bandwidth and 30.5 kHz frequency resolution. Twelve spectral windows were assigned to the $^{12}$CO ($J = 1$–0) rest frequency of 115.271202 GHz, and four to the $^{13}$CO ($J = 1$–0) rest frequency of 110.201353 GHz, both centered at the systemic velocity of DDO 154 at $V_{\rm lsr} = 374$ km s$^{-1}$. At 115 GHz, the frequency resolution corresponds to a velocity resolution of 0.08 km s$^{-1}$.

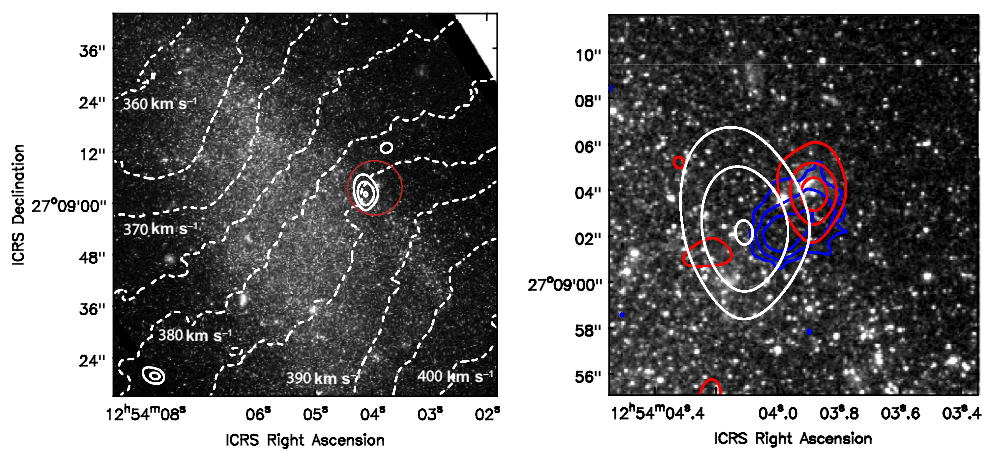

**Fig. 1.** Left: Stellar disk of DDO 154 as seen by the HST ACS *F606W*. White solid contours indicate cold dust traced by Herschel SPIRE 250 $\mu$m, drawn at 1.8, 2.0, and 2.2 MJy str$^{-1}$. Dashed lines indicate H I iso-velocity contours obtained at VLA (Walter et al. 2008), drawn at every 5 km s$^{-1}$. The red circle indicates the position of Region 2 and beam size (15″) of the NRO 45 m telescope observation. Right: Close-up on Region 2. Greyscale and white solid contours are the same as the figure on the left. Red contours indicate MIPS 24 $\mu$m, contours drawn at 0.1, 0.15, and 0.2 MJy str$^{-1}$. Blue contours indicate H$\alpha$ emission (Kennicutt et al. 2008).

Absolute intensity calibration was done using the standard chopper-wheel method. System temperatures varied from ∼250 to 400 K during the observations. IRC +10216 was observed daily as the standard source, and its integrated intensity varied by 13% over the observations, which we take to be the relative calibration uncertainty. The image rejection ratio (IRR) of the 2SB FOREST receiver was measured every day for each beam and polarization. For the $^{12}$CO and $^{13}$CO setups, the IRR was approximately 13–15 dB (corresponding to 3%–5%) and 15–20 dB (1%–3%), respectively, throughout the observing run. Thus, we ignore the effect of the IRR on the calibration uncertainty.

Pointing of the telescope was checked every 1–2 hr using nearby SiO masers, and any data with a pointing error larger than 5″ were discarded. The average pointing offset was 3″.5. Assuming the instantaneous pointing offset at a given time within the observation to be distributed normally, we can estimate the effective beam size by adding the HPBW of the telescope at 115 GHz (14″.5) and the pointing offset in quadrature, which gives 15″. We use this effective beam size throughout the paper.

Data reduction was done using NEWSTAR (Ikeda et al. 2001). The spectra were first inspected to flag out scans with spurious channels and baseline ripples. First-order baselines were fitted to the velocity ranges of 300 to 350 km s$^{-1}$ and 400 to 450 km s$^{-1}$, subtracted, averaged and smoothed to a velocity resolution of 1, 2, and 3 km s$^{-1}$. The antenna temperature $T_{\rm a}^*$ was converted to main beam temperature $T_{\rm mb}$ using $T_{\rm mb} = T_{\rm a}^*/\eta_{\rm mb}$, where $\eta_{\rm mb} = 0.39$ is the main beam efficiency for this observing run, provided by the observatory.

Figure 2 shows the resulting spectra at the $^{12}$CO ($J = 1$–0) and $^{13}$CO ($J = 1$–0) frequencies for a range of velocity binnings. The rms noise level in the velocity range 300 to 450 km s$^{-1}$ are shown as well.

The speculative $^{12}$CO signal of DDO 154 by K11 was 6 mK when observed with a 22″ IRAM 30 m beam and velocity binning of 3 km s$^{-1}$. This converts to an expected point source signal of ∼13 mK using the NRO beam. We do not detect any significant signal in any velocity binning. Similarly, no signals with $S/N > 3$ are found for the $^{13}$CO line.

## 3 Discussion

### 3.1 Upper limit on CO intensity

No signal was detected in $^{12}$CO towards Region 2. The $3\sigma$ upper limit on the CO intensity can be calculated by

$$\frac{I_{\rm CO}}{[\rm K\,km\,s^{-1}]} < 3 \frac{\sigma_{\rm rms}}{[\rm K]} \sqrt{\frac{\delta V}{[\rm km\,s^{-1}]} \frac{\Delta V}{[\rm km\,s^{-1}]}}, \quad (1)$$

where $\delta V$ and $\Delta V$ are the velocity bin and the line width, respectively. The non-detection does not allow for a measurement of $\Delta V$. K11 assumed a velocity width of 10 km s$^{-1}$, but in light of recent interferometric CO detections at similar metallicity (Rubio et al. 2015; Shi et al. 2020) which have found that CO line-widths are smaller than 2 km s$^{-1}$, we take $\Delta V = 2$ km s$^{-1}$ as a conservative value. The CO intensity upper limits are shown in figure 2.

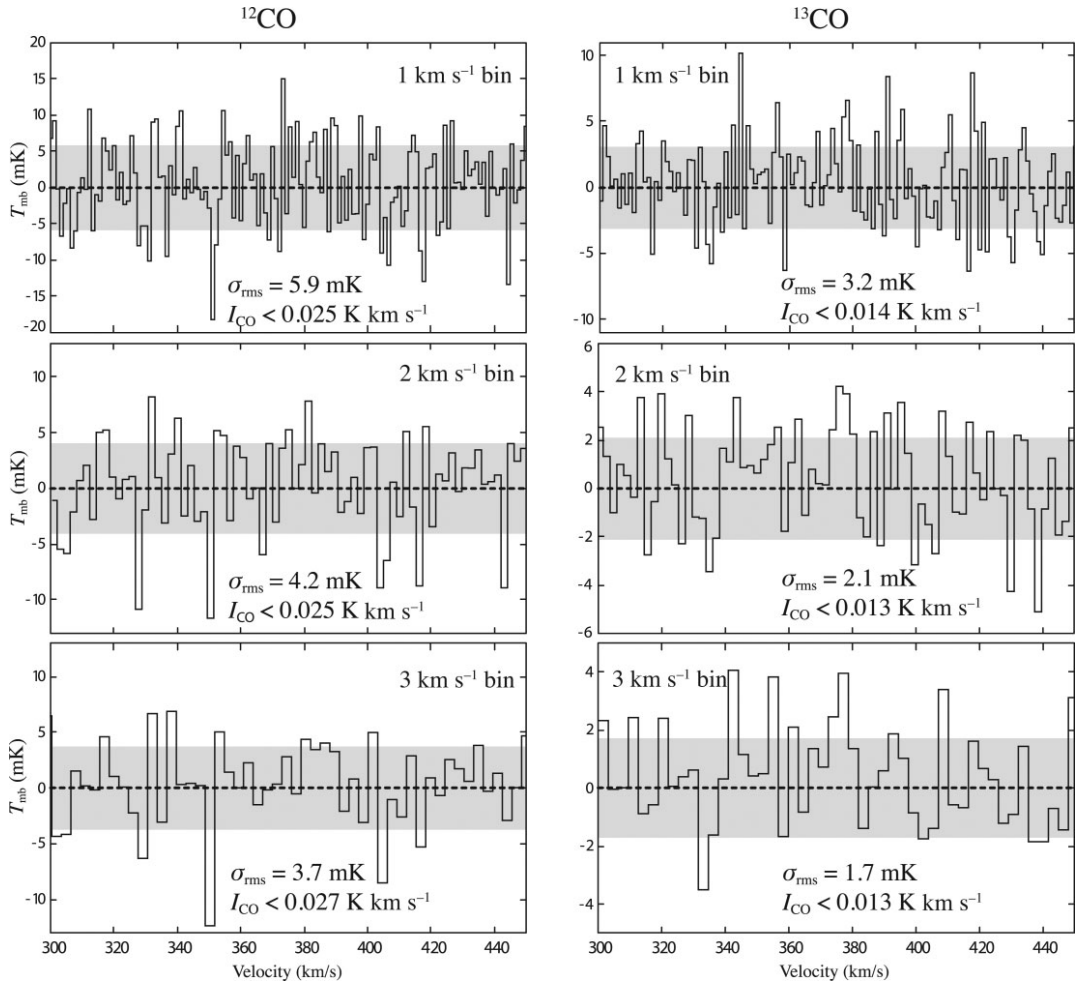

**Fig. 2.** Spectra of Region 2 at the $^{12}$CO (left) and $^{13}$CO (right) frequencies, for velocity binning of 1, 2, and 3 km s$^{-1}$. The shaded region corresponds to range within $\pm\sigma_{\rm rms}$.

The upper limit on CO can also be expressed in terms of luminosity, by

$$\frac{L'_{\rm CO}}{[{\rm K\,km\,s^{-1}\,pc^2}]} = 23.5 \frac{\Omega}{[{\rm arcsec}^2]} \left(\frac{D}{[{\rm Mpc}]}\right)^2$$
$$\times \frac{I_{\rm CO}}{[{\rm K\,km\,s^{-1}}]} < 1.06 \times 10^3, \quad (2)$$

where $\Omega$ is the solid angle of the 15″ beam and $D = 3.2$ Mpc is the distance to DDO 154, and we substituted the CO intensity upper limits for the 2 km s$^{-1}$ binned spectrum in the inequality.

We observed the $^{13}$CO frequency for the first time towards Region 2. For velocity bins of 2 and 3 km s$^{-1}$, two consecutive channels give S/N > 2 at the velocity of ∼380 km s$^{-1}$. For this signal to be a real detection, the $^{12}$CO/$^{13}$CO intensity ratio must be smaller than 1. Previous millimeter observations of the Galaxy and nearby galaxies have obtained $^{12}$CO/$^{13}$CO intensity ratios of >10 (Paglione et al. 2001; Davis 2014; Israel 2020). Simulations of turbulent molecular clouds by Szűcs, Glover, and Klessen (2014) have shown that isotope-selective photodissociation is not significant within a range of metallicities and field strengths, always giving $^{12}$CO/$^{13}$CO > 10. Thus, it is very unlikely that the ∼380 km s$^{-1}$ feature in figure 2 is actual $^{13}$CO emission.

### 3.2 Estimation of molecular gas mass

The molecular gas column density in Region 2 can be estimated using the SFR of Region 2 (Verter & Hodge 1995; Taylor & Klein 2001):

$$\frac{N({\rm H}_2)}{[{\rm cm}^{-2}]} = \frac{9.5 \times 10^{-23}}{\varepsilon_{\rm SF}} \frac{L({\rm H}\alpha)}{[{\rm erg\,s^{-1}}]} \frac{\tau_{\rm SF}}{[{\rm yr}]} \left(\frac{[{\rm pc}]}{R}\right)^2, \quad (3)$$

where the local H$\alpha$ luminosity is $L({\rm H}\alpha) = 7.0 \times 10^{36}$ erg s$^{-1}$ (Kennicutt & Skillman 2001). K11 assume the star formation timescale $\tau_{\rm SF}$ to be $10^8$ yr, which is limited by the dynamical timescale of DDO 154 and the disruption of molecular gas by OB stars (Egusa et al. 2004, 2009). The star formation efficiency $\epsilon_{SF}$ is assumed to be 0.2 (Yasui et al. 2008). The cloud size $R$ was assumed to be 50 pc, but since we assume in subsection 3.1 that the cloud has a velocity width of $\Delta V = 2$ km s$^{-1}$, we change the cloud size accordingly. All detected clouds in WLM have radii of less than 6 pc (Rubio et al. 2015), and those in Sextans B have radii less than 3 pc (Shi et al. 2020), both roughly following the classical size–line width relation Solomon et al. (1987). Here we assume that the cloud has a radius of 5 pc. Substituting all of these values into equation (3), we estimate $N(H_2) = 1.3 \times 10^{22}$ cm$^{-2}$. The corresponding molecular cloud mass is $M(H_2) = 8.3 \times 10^3 M_\odot$.

An alternative estimate of the molecular gas mass can be obtained by using cold dust as a proxy of atomic and molecular gas. The molecular gas mass, atomic gas mass $M(H_I)$, and dust mass $M_d$ are connected via the gas-to-dust ratio $\delta_{GDR}$, so that

$$\delta_{GDR} = \frac{M(H_I) + M(H_2)}{M_d} \times 1.36, \quad (4)$$

where the factor 1.36 accounts for the contribution of helium, and here we assume that gas and dust are well mixed. We can estimate $M_d$ from Herschel SPIRE 250 $\mu$m photometry from the KINGFISH project (Kennicutt et al. 2011), indicated as white contours in figure 1. The flux within the NRO beam is $S_d = 19.1$ mJy, after applying an aperture correction factor of 3.3 applicable to a point source measured in an aperture with radius of 7″ (Smith et al. 2017). Assuming that the dust emissivity does not change for the atomic and molecular phases, the dust mass is calculated by

$$\frac{M_d}{[M_\odot]} = 4.79 \times 10^{14} \frac{S_d}{[Jy]} \left(\frac{D}{[Mpc]}\right)^2 \frac{[m^2 \, kg^{-1}]}{\kappa_d} \frac{[Jy]}{B_\nu(T)} \quad (5)$$

where $D = 3.2$ Mpc is the distance, $\kappa_d$ is the dust emissivity coefficient at 250 $\mu$m, and $B_\nu(T)$ is the Planck function at dust temperature $T$. Various values of $\kappa_d$ can be found in the literature. Bianchi et al. (2019) show that DustPedia galaxies with low metallicity (12 + log[O/H] $\sim$ 8.0) have $\kappa_d$ ranging from 0.1 to 0.3 m$^2$ kg$^{-1}$ but with a limited number of samples. Larger values like 0.4 m$^2$ kg$^{-1}$ (Bianchi 2013) and 0.48 m$^2$ kg$^{-1}$ (Dale et al. 2012) can be found for a statistical sample, depending on the data used and how the SED is modeled. Here we take $\kappa_d = 0.3$ m$^2$ kg$^{-1}$ as a rough estimate, and assume an uncertainty of 50% which would effectively enclose the variations in literature.

At 250 $\mu$m, dust is heated by the ambient interstellar radiation field (Planck Collaboration 2014) and its temperature in nearby galaxies is normally around $T \sim 20$ K

(Komugi et al. 2011a; Galametz et al. 2012). In dwarf irregular galaxies with extremely low metallicity, Zhou et al. (2016) use two-temperature modified blackbody fits to Herschel photometry and obtain cold dust temperatures ranging from 13 to 20 K in the range 7.2 < 12 + log[O/H] < 7.9. Chang et al. (2021) use Bayesian fitting to the dwarf irregular UGC 04305 with 12 + log[O/H] = 7.77 and obtain $T = 12$ K. Here we assume a cold dust temperature of 15 K, and assign an uncertainty of ±5 K, or $\sim$30%. Using these values and adding the errors in quadrature, we estimate $M_d = (5.6 \pm 3.4) \times 10^3 M_\odot$.

Atomic gas (H I) was measured by VLA THINGS (Walter et al. 2008). The flux within the 15″ NRO beam was $S_{HI} = 0.33$ Jy km s$^{-1}$. We determine the H I mass using

$$\frac{M(H_I)}{[M_\odot]} = 2.356 \times 10^5 \frac{S_{HI}}{[Jy \, km \, s^{-1}]} \left(\frac{D}{[Mpc]}\right)^2 \quad (6)$$

and obtained $M(H_I) = 7.9 \times 10^5 M_\odot$. Photometry errors were negligible.

The gas-to-dust ratio $\delta_{GDR}$ is known to correlate with the metallicity. Sandstrom et al. (2013) measure $\delta_{GDR}$ in nearby galaxies as a function of metallicity. Substituting the metallicity of DDO 154 into their relation, we obtain $300 < \delta_{GDR} < 1150$, where the range of ratio comes from different metallicity calibrations (Kobulnicky & Kewley 2004; Pilyugin & Thuan 2005). Magdis et al. (2012) find $\log \delta_{GDR} = 10.54 - 0.99(12 + \log[O/H])$ with a scatter of 0.15 dex for local and $0.5 < z < 2$ galaxies combined, which gives $\delta_{GDR} = 880$ for DDO 154. An important caveat is that these relations are constructed with galaxies in the range $8 < 12 + \log[O/H]$, and estimated values at the metallicity of DDO 154 are extrapolations. Here we take $\delta_{GDR} = 880$ from Magdis et al. (2012) as a fiducial value, and allow for a factor 2 uncertainty.

Substituting these estimates into equation (4), we obtain the dust-based molecular mass estimate of $M(H_2) = 2.8 \times 10^6 M_\odot$. If we take the range of $M_d$ within the uncertainties and further allow for a factor 2 change in $\delta_{GDR}$, the molecular gas mass can range from $6.3 \times 10^5 M_\odot$ to $8.7 \times 10^6 M_\odot$.

It is important to point out that the two commonly used methods to estimate molecular gas mass, one based on local SF arguments [$M(H_2) = 8.3 \times 10^3 M_\odot$] and the other based on dust [$M(H_2) = 2.8 \times 10^6 M_\odot$], are inconsistent by two orders of magnitude. Realistic variations in $\tau_{SF}$, $\epsilon_{SF}$ in equation (3) and/or $\delta_{GDR}$, $\kappa_d$, and $T$ cannot bring the two estimates to within uncertainties. A possible way to reconcile these differences is the coefficient to the H$\alpha$ luminosity used in equation (3). Equation (3) assumes that the H$\alpha$ luminosity can be converted linearly to the local SFR, based on a Salpeter initial mass function (Salpeter 1955;

Kennicutt 1983). At low SFR, however, the H$\alpha$ luminosity may severely underestimate the true SFR, as the initial mass function may be undersampled (Pflamm-Altenburg et al. 2007). For Region 2, the observed H$\alpha$ luminosity may underestimate the SFR by 1.5–2.5 orders of magnitude, depending on the initial mass function used. Here we do not take this effect into account, as the Verter and Hodge (1995) relation in equation (3) will give a conservative estimate of the molecular gas mass, and thus a lower limit for the conversion factor.

### 3.3 Conversion factor

Using the upper limit on CO intensity $I_{CO}$ derived in subsection 3.1 and the molecular gas column density $N(H_2)$ in subsection 3.2, we can estimate the lower limit on the conversion factor $X_{CO}$ by

$$\frac{X_{CO}}{[\text{cm}^{-2}(\text{K km s}^{-1})^{-1}]} = \frac{N(H_2)}{I_{CO}} > 5.2 \times 10^{23}, \quad (7)$$

which is three orders of magnitude larger than the typical MW value. This is consistent with previous estimates (Wilson 1995; Arimoto et al. 1996; Israel 1997; Leroy et al. 2011; Bolatto et al. 2013; Hunt et al. 2023) extrapolated to low metallicity, but much higher than recent estimates based on actual CO detection with interferometers in dwarf irregular galaxies with similar metallicity, which give $X_{CO}$ values of only several times the MW value (Rubio et al. 2015; Shi et al. 2020). These estimates based on interferometric CO detection assume that CO is virialized, and indicates the conversion factor limited to the region where CO is detected. Since our estimate of $N(H_2)$ comes from gas directly connected to local SF, and thus naturally includes the total gas outside of where CO is detected, the high lower limit on $X_{CO}$ is expected. The high limit also comes from the revised assumption of the cloud size and velocity width in light of recent results. If we assume the same molecular cloud parameters as in K11 with CO velocity width of 10 km s$^{-1}$ and size of 50 pc, the lower limit becomes $X_{CO} > 2.6 \times 10^{21}$, consistent with the lower limit estimate in K11. A large CO-dark molecular gas fraction is consistent with the flattening of the H I radial profile in the central region of DDO 154 (Hunter et al. 2021).

The conversion factor $\alpha_{CO}$ corresponding to the dust-based estimate of molecular gas $M(H_2)$ is related to the CO luminosity by

$$\frac{\alpha_{CO}}{[M_\odot(\text{K km s}^{-1}\text{ pc}^2)^{-1}]} = \frac{M(H_2)}{L'_{CO}}. \quad (8)$$

Using equation (8) and the CO luminosity lower limit derived in subsection 3.1, we obtain the dust-based conversion factor $\alpha_{CO} > 2.7 \times 10^3$, which is $\sim$600 times larger than the canonical value in the MW, $\alpha_{CO, MW} = 4.4$ (Bolatto et al. 2013). This lower limit on the dust-based conversion factor is crude at best, however. Accounting for the range of dust-based molecular mass, the lower limit on $\alpha_{CO}$ can be as low as 600, which is still two orders of magnitude larger than the MW. The result is consistent with other dust-based estimates of the conversion factor (Leroy et al. 2011; Sandstrom et al. 2013) in nearby galaxies, extrapolated to extreme low metallicity.

### Acknowledgments

This work was supported by JSPS KAKENHI Grant Number 20K04015. This paper is based on observations at the Nobeyama Radio Observatory (NRO) 45 m radio telescope. The Nobeyama 45 m radio telescope is operated by Nobeyama Radio Observatory, a branch of National Astronomical Observatory of Japan.